\documentclass{aa}  
\usepackage{graphicx}
\usepackage{txfonts}
\usepackage{lipsum}
\usepackage{subcaption}         % necessary for continued figures, example in section 3
\usepackage{lscape}             % to rotate a single page table, example in appendix.
\usepackage{placeins}           % useful with \FloatBarrier, to keep 
\usepackage{natbib}
\usepackage{graphicx}	% Including figure files
\usepackage{amsmath}	% Advanced maths commands
\usepackage{amstext,amscd,bm}
\usepackage{array}
\usepackage[colorlinks=true,
citecolor=blue,
linkcolor=blue,
urlcolor=blue]{hyperref}
                               
\begin{document}

%%%%%%%%%%%%%%%%%%%%%%%%%%%%%%%%%%%%%%%%
% if you use custom commands in your title,
% ensure to check your title when submitting!
%%%%%%%%%%%%%%%%%%%%%%%%%%%%%%%%%%%%%%%%
   \title{No Convincing Evidence for Tidal Binary Clusters in the Galaxy }

   %\subtitle{Subtitle}

%%%%%%%%%%%%%%%%%%%%%%%%%%%%%%%%%%%%%%%%
% Please separate each author with the \and command
%
% Use the \corrauth to provide the corresponding
% author address. It will be automatically inserted as 
% footnote in the PDF output.
%
% Please DO NOT include ORCIDs next to author names.
% Instead, please provide an active address for each coauthor:
% it will be automatically extracted by EDPS editorial system, 
% and co-authors will be be able to authenticate their ORCID.
%
% Only authenticated ORCIDs will be taken into account.
% ORCIDs included here will be removed.
%%%%%%%%%%%%%%%%%%%%%%%%%%%%%%%%%%%%%%%%

   \author{Juan Casado\inst{1}\corrauth{jcasadoo@hotmail.com}        % use \corrauth for the corresponding author
        \and Zahra Al\inst{2}\email{zahraa96al@gmail.com}
        }

    \institute{Facultad de Ciencias, Universitat Autònoma de Barcelona (UAB), 08193 Bellaterra, Barcelona, Spain
	\and Institute of Graduate Studies In Sciences, {\.I}stanbul University, 34116, \"Universite-Istanbul, T\"urkiye}

   \date{Received September 30, 2026}

% \abstract{}{}{}{}{}
% 5 {} token are mandatory
 
  \abstract
% context heading (optional)
{The existence of tidal binary clusters (TBCs) in the Milky Way has been repeatedly suggested using Gaia DR3 astrometry and recent catalogs of binary clusters and groups. However, the dynamical feasibility of forming truly bound binary clusters through tidal capture remains uncertain.}
% aims heading (mandatory)
{We perform a systematic reassessment of twenty star clusters, currently claimed as members of TBCs from the catalog of binary and grouped clusters of \cite{liu25} to ascertain their classification.} 
% methods heading (mandatory)
{For each candidate pair, we use Gaia DR3 astrometry and photometry to perform a simultaneous and self-consistent optimization of age, distance modulus, reddening, and metallicity.}
% results heading (mandatory)
{Our analysis shows that all studied TBCs fail in the required physical condition that ages are compatible with independent formation. After applying strict evolutionary criteria, these systems are more naturally explained as co-eval members of primordial cluster pairs or groups rather than products of tidal capture.}
% conclusions heading (optional)
{We conclude that genuine tidal binary clusters, if present in the Milky Way, must be extremely rare. This work underscores the importance of robust age determinations and physically motivated selection criteria in future searches for dynamically assembled cluster pairs.}


   \maketitle

%%%%%%%%%%%%%%%%%%%%%%%%%%%%%%%%%%%%%%%%%%%%%%%%%%%%%%%%%%%%%%
\section{Introduction}

\nolinenumbers

Open clusters originate within giant molecular clouds (GMCs) \citep{lada2003}, where hierarchical fragmentation can produce not only isolated systems but also cluster pairs and larger primordial groups. In such a formation scenario, double or multiple clusters emerge nearly simultaneously from the same parental cloud and are therefore expected to share similar ages, chemical compositions, distances, and kinematics \citep{deoliveira2000, conrad17, casado22b, song22}. Observational studies of the Magellanic Clouds and the Milky Way support this scenario: young binary clusters often display nearly identical color–magnitude diagrams and common proper motions, consistent with a shared origin through GMC fragmentation \citep{deoliveira2000, diabel2002, song22}. 

Alternative mechanisms have also been proposed to explain the formation of cluster pairs. In sequential formation, stellar winds and supernova feedback from a young cluster compress nearby gas, triggering the birth of a companion cluster \citep{palma25, li25}. A more controversial pathway is tidal capture, in which two independently formed clusters become gravitationally bound after a close encounter. During such an interaction, tidal forces might dissipate orbital energy and angular momentum, potentially leaving the clusters in a bound configuration \citep{palma25, li25}. Because these systems do not share a common origin, they are expected to exhibit markedly different ages and chemical compositions. In this work, we use the term tidal binary clusters (TBCs) specifically to refer to cluster pairs formed via tidal capture. 

The physical nature of candidate TBCs remains highly uncertain. Spatial proximity and similar proper motions alone do not grant gravitational binding, and recent studies increasingly suggest that many proposed systems are primordial or transient flyby encounters rather than genuine binaries \citep{casado22a, liu25}. \cite{casado22a} found no convincing evidence for a true binary among 120 revised pairs or groups comprising at least one cluster older than 100 Myr, leaving only two ambiguous cases. These results reinforce the view that binary open clusters are dynamically fragile and survive only for relatively short periods. 

This perspective is formalized in the Primordial Group hypothesis \citep{casado22b}, which argues that only sufficiently young clusters can exist in binary or multiple configurations. Old open clusters are expected to be essentially single. According to this hypothesis, genuinely old binary clusters should not exist in the Milky Way. Cluster groups are therefore interpreted as a transient, primordial phenomenon. 

Theoretical studies support this interpretation. Numerical simulations show that binary clusters are dynamically unstable, with typical survival timescales of only 10–200 Myr depending on separation, mass ratio, and Galactic environment \citep{portegies07, priyatikanto16}. Pairs born too close would rapidly merge due to angular momentum loss, whereas wider pairs are easily disrupted by Galactic tides. The Milky Way disk, dominated by strong tidal shear, differential rotation, and frequent encounters with GMCs, is particularly hostile to weakly bound systems \citep{berg96, Marcos09, priyatikanto16}. In binary systems, the combined tidal tensor of the Galaxy and the companion cluster produce asymmetric distortions and tidal bridges that accelerate disaggregation. 

Simulations of close encounters further support the transient nature of binary clusters. \cite{Zhu2026} show that capture or merger events occur only under extremely restrictive conditions, requiring nearly parabolic encounters with substantial overlap of their stellar envelopes. 

Despite this, claims of old binary clusters continue to appear. \cite{camargo21} proposed NGC 1605 as an old binary system, reporting two apparent cores separated by ~1.8 pc and with ages differing by ~1.4 Gyr. However, subsequent studies using Gaia EDR3 astrometry \citep{andres22} and CCD UBV plus Gaia DR3 data \citep{gokmen23} found no evidence supporting the existence of a second physical cluster. The NGC 1605 case remains notable because old binary-cluster candidates are exceedingly rare. 

Recent Gaia DR3 studies have identified many candidate binary or multiple systems. \cite{palma25} reported 617 paired systems and 261 larger groups using tidal factors and nearest-neighbor statistics. \cite{liu25}, using a high-quality sample of 4,084 clusters, identified 400 pairs and 82 multiple systems. Both studies suggest that tidal interactions may be common during cluster evolution and can lead to TBCs. However, a detailed follow-up analysis challenges this interpretation. \cite{zhu25} examined two candidate TBCs—ASCC 71 \& ESO 064-05 and NGC 2129 \& UBC 437—previously classified as capture systems. Although both pairs exhibit significant age differences consistent with independent formation, orbital integrations and N-body simulations show that neither system is gravitationally bound, and both are expected to separate rapidly. These results demonstrate that present-day proximity alone is insufficient to identify genuine tidal-capture binaries. Moreover, tidal capture itself is expected to be extremely rare in the Milky Way due to its relatively high stellar velocity dispersion. 

In this work, we continue to examine the Primordial Group hypothesis by reassessing the evidence for TBCs in the Galaxy using the catalog of \cite{liu25}. Section 2 describes data selection and methodology. Section 3 presents the results and discusses them. In Section 4, we summarize the main conclusions and outline future research directions. 

% %_____________________________________________________________
% %                      Another onecolumn float (place early!)
% %-------------------------------------------------------------
% \begin{figure*}[ht!]
%    \resizebox{\hsize}{!}
%             {\includegraphics {figure.eps}}
%             %{\includegraphics[bb=10 20 100 100,clip]{figure.eps}}
%       \caption{Another onecolumn figure*}
%          \label{wide}
% \end{figure*}
% \twocolumn
% \clearpage
% % %%%%%%%%%%%%%%%%%%%%%%%%%%%%%%%%%%%%%%%%%%%%%%%%%%%%%%%%%%%%%%

\section{Method}\label{method}
\subsection{Data Selection}\label{sec:data_Sel}

Our initial sample was drawn from the binary cluster catalog of \citet{liu25}, which reports several hundred candidate physical pairs, including numerous newly identified systems. From this catalog, we selected all objects classified as tidal binary clusters (TBCs), yielding a total of 292 pairs. Within this subset, we restricted our analysis to systems with substantial age differences, as such discrepancies are expected for most clusters that formed independently and were subsequently captured dynamically. 

To identify the most promising candidates, we enforced two strict selection criteria:
\begin{enumerate}[(i)]
	\item At least one component must have an age $\ge 300~\text{Myr}$, thereby excluding, in principle, a primordial co-eval origin within a common giant molecular cloud (GMC).
	\item The age difference between the two constituent clusters must exceed $100~\text{Myr}$, to avoid sequential starburst episodes within the same parental cloud.
\end{enumerate}
These initial cuts reduced the sample to 23 candidate systems. 

Given the well-known systematic uncertainties regarding open cluster age determinations, we cross-checked the ages of these 23 pairs using two independent sources: \citet{liu25} and \citet{H24}. Several systems don't satisfy both selection criteria when evaluated across both catalogs. Retaining those pairs that strictly satisfy both selection thresholds in both datasets resulted in a refined sample of 9 TBC candidates (18 unique clusters). The primary astrophysical properties of these systems are summarized in Table~\ref{tab:sample}. 

Furthermore, for Pairs 58 and 283, we incorporated HSC 1828 and HSC 1843, respectively, into our Color-Magnitude Diagram (CMD) analysis, following \citet{palma25}, who classify these additional objects as confirmed members of triple cluster systems. With these additions, the total number of individual clusters examined in this work is 20.

\subsection{Data Handling}

For metallicity (\textit{Z}), both \cite{H24} and \cite{liu25} adopted solar values. However, imposing a fixed solar metallicity for all clusters may introduce systematic biases in the determination of other astrophysical parameters. To avoid such biases, we treated \textit{Z} as a free parameter and performed a simultaneous optimization of all relevant system properties. 

The optimization was carried out using Gaia photometry for the member stars numbered in Table~\ref{tab:sample}. For each TBC, we first inspected the CMDs of the two clusters to evaluate whether they could be described by a single isochrone, as expected for systems sharing a common evolutionary history. For all systems, the CMDs were consistent with a shared isochrone. Thus, we handled both clusters simultaneously to derive their best‑fitting parameters. 

Following the approach of \citet{al2026}, the fitting procedure employed the \texttt{simul\_fit} code\footnote{\url{https://github.com/speletry/simul_fit.py}}, a Python tool designed for multi-filter, simultaneous isochrone fitting based on PARSEC evolutionary models \citep{bressan12}\footnote{\url{https://stev.oapd.inaf.it/cmd/}}. The fits were performed using Gaia $G$ versus $(G_{\text{BP}} - G_{\text{RP}})$ CMDs, and the optimal solution was obtained by minimizing the global $\chi^{2}$. In this way, the parameters of metallicity ($Z$), distance modulus ($\text{DM}$), reddening $E(B-V)$, and age were jointly determined.

Formal $1\sigma$ uncertainties and parameter covariances were computed from the $\chi^{2}$ grid using likelihood weights of the form 

\begin{equation}
	P \propto \exp \left( -\frac{\Delta \chi^2}{2} \right)
\end{equation}

\begin{table*}
	\caption{ Selected open cluster pairs from Liu et al. 2025 based on criteria given in Sec.~\ref{sec:data_Sel}. Ages from \citet{liu25} and \citet{H24} are provided in Myr columns 5 and 7, respectively.}
	\label{tab:sample}
	\centering
	\begin{tabular}{llccccc}
		\hline\hline
		\noalign{\smallskip}
		Pair & Clusters & Members & $\log(t/\text{yr})$ & Age & $\log(t/\text{yr})$ & Age \\
		& & & Liu25 & (Myr) & HR24 & (Myr) \\
		\noalign{\smallskip}
		\hline
		\noalign{\smallskip}
		15  & ASCC 108     & 305 & 8.05 & 112 & 8.28 & 191 \\
		& HSC 582      & 39  & 8.90 & 794 & 8.75 & 562 \\
		\noalign{\smallskip}
		58  & CWNU 110     & 40  & 8.07 & 117 & 8.34 & 219 \\
		& NGC 2428     & 274 & 8.72 & 525 & 8.97 & 933 \\
		\noalign{\smallskip}
		201 & UBC 1320     & 29  & 8.19 & 155 & 8.33 & 214 \\
		& HSC 1585     & 28  & 8.63 & 427 & 8.72 & 525 \\
		\noalign{\smallskip}
		259 & HXHWL 65     & 55  & 8.31 & 204 & 8.68 & 479 \\
		& Ruprecht 53  & 70  & 8.68 & 479 & 8.54 & 347 \\
		\noalign{\smallskip}
		283 & Theia 366    & 81  & 8.00 & 100 & 8.22 & 166 \\
		& NGC 2358     & 169 & 8.55 & 355 & 8.62 & 417 \\
		\noalign{\smallskip}
		329 & Theia 236    & 16  & 6.92 & 8   & 7.90 & 79  \\
		& PHOC 40      & 75  & 8.61 & 407 & 8.57 & 372 \\
		\noalign{\smallskip}
		343 & Teutsch 247  & 182 & 7.74 & 55  & 7.76 & 58  \\
		& Trumpler 23  & 650 & 8.61 & 407 & 8.52 & 331 \\
		\noalign{\smallskip}
		361 & Theia 2151   & 33  & 7.89 & 78  & 8.40 & 251 \\
		& UBC 502      & 86  & 8.49 & 309 & 8.50 & 316 \\
		\noalign{\smallskip}
		369 & Theia 4284   & 20  & 8.81 & 646 & 7.94 & 87  \\
		& Theia 4981   & 65  & 8.67 & 468 & 8.54 & 347 \\
		\noalign{\smallskip}
		\hline
	\end{tabular}
\end{table*}

\section{Results and Discussion}\label{sec:results}

\begin{figure*}
	\centering
	\includegraphics[width=0.96\textwidth]{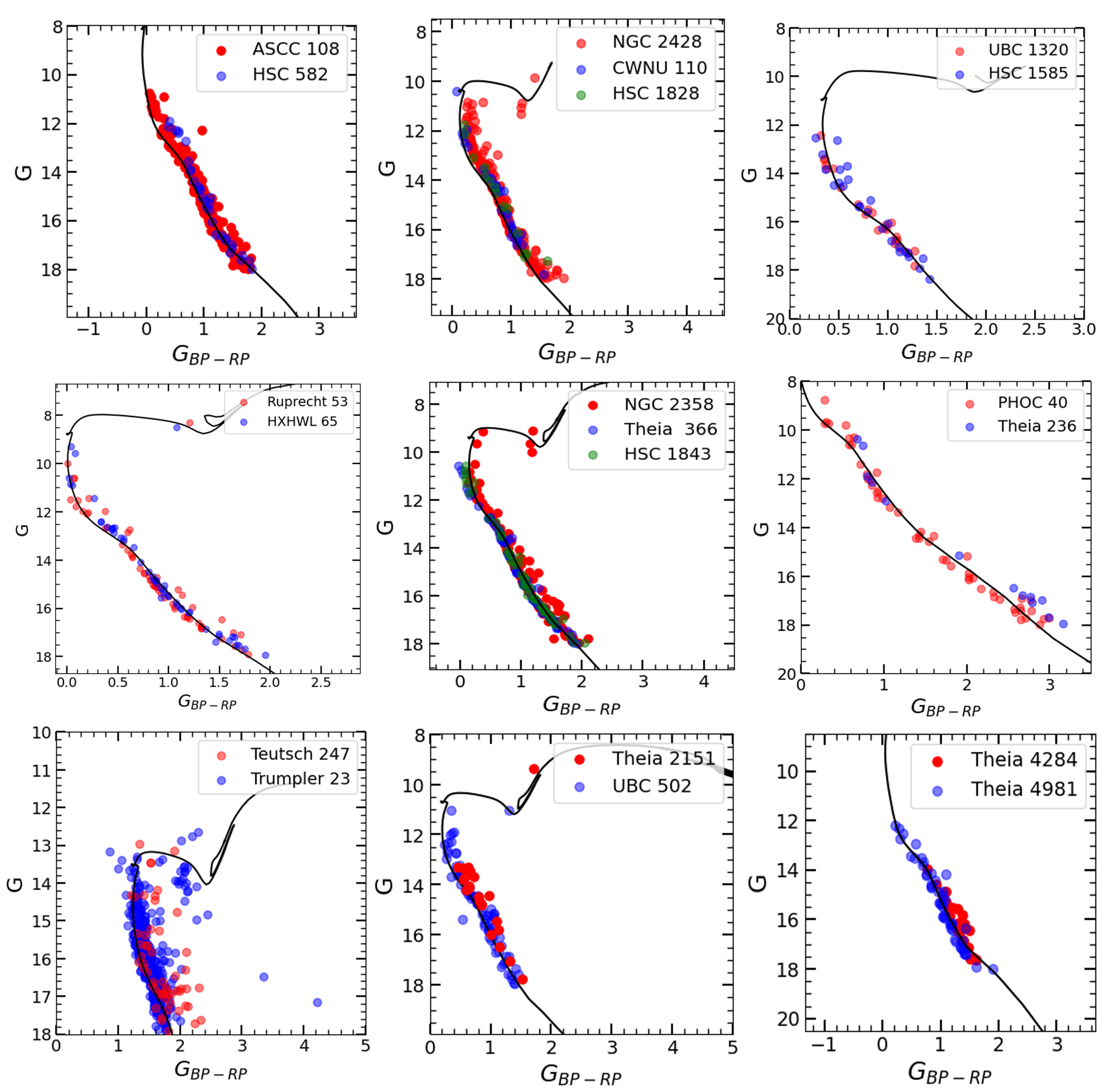}
	\caption{$G$ versus $(G_{\text{BP}} - G_{\text{RP}})$ CMDs of the nine TBCs candidates studied. The CMDs were constructed using the member stars identified for each cluster by \citet{H24}. For pairs 58 and 283, a third cluster identified as a member of the system by \citet{palma25} is also included in the CMD. As shown in the nine CMDs, the clusters in each system are successfully described by a single isochrone, supporting their coeval nature.}	
	\label{fig:cmd_fits}
\end{figure*}

Each group features overlapping distributions in their color-magnitude diagrams (CMDs) that can be well characterized using a single isochrone. The resulting isochrone fits to the Gaia $G$ versus $(G_{\text{BP}} - G_{\text{RP}})$ CMDs are presented in Fig.~\ref{fig:cmd_fits}, and the derived best-fit parameters are summarized in Table~\ref{tab:best_fit_params}, together with the corresponding reduced chi-squared values ($\chi^2_{\text{red}}$), which range from 0.66 to 0.88. These values, being close to unity, indicate that the adopted isochrones provide an accurate representation of the observed CMD distributions.

Treating metallicity ($Z$) as a free parameter yields individualized values that generally scatter within the expected Galactic ranges. Many of the systems analyzed here lack detailed spectroscopic evaluations in the literature, leading to broad variations in \textit{Z} through comparative catalogs. The remaining fundamental parameters in Table~\ref{tab:best_fit_params} are broadly consistent with the values reported in the  literature.

Focusing on the primary objective of this study -evaluating cluster coevality- the ages derived here for Pairs 15, 201, 259, and 369 are systematically lower than those listed in Table~\ref{tab:sample}, whereas for the remaining pairs, our derived ages fall within the range of individual values reported for the two constituent clusters. Four out of the nine systems yield ages close to $300~\text{Myr}$ (probably a consequence of our initial selection criteria targeting relatively evolved clusters), suggesting that this threshold represents a practical upper age limit for primordial cluster groups. The results depicted in Fig.~\ref{fig:cmd_fits} strongly indicate that all studied groups are coeval. Consequently, the large age discrepancies between components of the same pair reported in earlier literature (Table~\ref{tab:sample}) are likely spurious and do not provide compelling evidence for a tidal-capture origin. In several instances, we have found additional evidence that supports the case for primordial groups or, alternatively, challenges the case for TBCs (see Appendix A). In particular, there are significant inconsistencies in the classification of binary clusters in different catalogs. For example,  \citet{palma25} identify most of our sample as mere optical pairs, which result from hyperbolic encounters but are not gravitationally bound.

\begin{table*}
	\caption{Astrophysical parameters derived from the best-fitting isochrones obtained with the simultaneous fitting procedure for the nine candidate tidal binary pairs. The listed parameters are respectivelly the color excess $E(B-V)$, distance modulus ${DM}$, heliocentric distance $D$, logarithmic age $\log(t/\text{yr})$, age (Myr), metallicity $Z$, and  $\chi^2_{\text{red}}$ of the fits.}
	\label{tab:best_fit_params}
	\centering
	\begin{tabular}{lccccccc}
		\hline\hline
		\noalign{\smallskip}
		Pair & $E(B-V)$ & ${DM}$ & $D$ (kpc) & $\log(t/\text{yr})$ & Age (Myr) & $Z$ & $\chi^2_{\text{red}}$ \\
		\noalign{\smallskip}
		\hline
		\noalign{\smallskip}
		15  & $0.12 \pm 0.02$ & $10.2 \pm 0.1$ & $1.1 \pm 0.1$  & $7.9 \pm 0.1$ & 79  & $0.017 \pm 0.001$ & 0.68 \\
		58  & $0.15 \pm 0.02$ & $10.9 \pm 0.5$ & $1.5 \pm 0.35$ & $8.5 \pm 0.1$ & 316 & $0.017 \pm 0.001$ & 0.79 \\
		201 & $0.43 \pm 0.02$ & $12.3 \pm 0.6$ & $2.88 \pm 0.80$& $7.9 \pm 0.1$ & 79  & $0.021 \pm 0.001$ & 0.81 \\
		259 & $0.12 \pm 0.02$ & $10.2 \pm 0.4$ & $1.1 \pm 0.2$  & $8.2 \pm 0.1$ & 158 & $0.017 \pm 0.001$ & 0.88 \\
		283 & $0.17 \pm 0.02$ & $10.0 \pm 0.4$ & $1.0 \pm 0.18$ & $8.5 \pm 0.1$ & 316 & $0.015 \pm 0.001$ & 0.84 \\
		329 & $0.10 \pm 0.02$ & $7.4 \pm 0.1$  & $0.30 \pm 0.02$& $8.1 \pm 0.1$ & 126 & $0.020 \pm 0.001$ & 0.75 \\
		343 & $1.0 \pm 0.2$   & $12.0 \pm 0.5$ & $2.5 \pm 0.58$ & $8.5 \pm 0.1$ & 316 & $0.024 \pm 0.001$ & 0.78 \\
		361 & $0.20 \pm 0.02$ & $11.3 \pm 0.5$ & $1.8 \pm 0.42$ & $8.5 \pm 0.1$ & 316 & $0.022 \pm 0.001$ & 0.69 \\
		369 & $0.24 \pm 0.02$ & $10.4 \pm 0.2$ & $1.2 \pm 0.1$  & $7.8 \pm 0.1$ & 63  & $0.009 \pm 0.001$ & 0.66 \\
		\noalign{\smallskip}
		\hline
	\end{tabular}
\end{table*}

Binary clusters are more frequent in Magellanic Clouds \citep{Bhatia91, diabel2002}. More than 69 double clusters have been identified within the LMC with separations below approximately 18 pc \citep{Bhatia88, Groenewegen2003}. Tidal capture might be more plausible in such environments, where the velocity dispersion is lower. Compared with the Milky Way, weaker tidal forces are exerted in the Magellanic Clouds \citep{Darma21}. Some studies revealed distorted isophotes, tidal bridges, common envelopes, and tidal tails between cluster pairs in the SMC \citep{deoliveira2000}. Approximately 25\% of the observed systems show clear morphological evidence of ongoing gravitational interaction. However, these systems are also predominantly young and coeval, supporting primordial formation rather than tidal capture. Similar conclusions were reached by \cite{diabel2002}, who argued that capture events are statistically improbable and that most genuine binary clusters originate through fragmentation of the same molecular cloud complex. 

Although tidal capture of star clusters is one of the most likely explanations for the existence of diverse star populations within globular clusters \cite{berg96}, in the present-epoch Milky Way, most candidate pairs are young open clusters. This, together with the results of \cite{casado22a} and other dynamical studies, suggests that true tidal-capture binaries, if they exist at all, are exceedingly rare. Furthermore, confirming tidal capture remains challenging and requires particularly stringent tests:  parallax agreement within uncertainties, ages compatible with independent formation, and relative velocities not exceeding the escape velocity at the observed separation.  

\section{Summary and Conclusion}\label{conclusion}

In this work, we reassessed the evidence for tidal binary clusters (TBCs) in the Milky Way using the catalog of \citet{liu25} and a homogeneous re-evaluation of the astrophysical parameters of the proposed systems. Starting from 292 pairs classified as TBCs, we applied physically motivated criteria requiring (i) at least one cluster to be older than $300~\text{Myr}$, and (ii) an age difference larger than $100~\text{Myr}$, as expected for systems formed independently and subsequently captured dynamically. These thresholds reduced the sample to 23 candidates. Given the substantial uncertainties that affect the age determinations of the open clusters, we cross-match the ages of these systems using two independent catalogs \citep{liu25, H24}. Only nine pairs simultaneously satisfied the selection criteria in both datasets, forming our best sample of TBC candidates.

To refine the physical characterization of these systems, we performed simultaneous PARSEC isochrone \citep{bressan12} fitting using Gaia photometry and treating \textit{Z} as a free parameter. This approach avoids the systematic biases introduced when \textit{Z} is fixed to solar values, as done in the mentioned catalogs, and allows a consistent optimization of age, distance modulus, reddening, and \textit{Z}. We find that these large-gap candidate systems are well-characterized by common isochrones. The resulting parameter sets lead to the conclusion that many alleged TBCs are simply and naturally explained as remnants or parts of primordial groups rather than products of dynamical capture. 

Thus, although candidate TBCs have been proposed, there is currently no compelling evidence that a long-lived tidal-capture binary cluster exists in the Milky Way. The apparent age discrepancies reported across different catalog compilations likely reflect the inherent parameter degeneracies and varied methodological assumptions in large-scale automated fits, rather than physical age gaps resulting from independent capture events. 

Overall, current observational and theoretical evidence increasingly supports the view that most binary open clusters are primordial systems formed within the same GMC, whereas long-lived TBCs in the Milky Way, if they exist, are extremely rare. We have shown that strong candidate TBCs identified in Gaia-era catalogs may instead represent primordial, coeval pairs rather than truly tidal-capture pairs, although some transient flyby encounters cannot be ruled out. The apparent absence of TBCs strongly suggests that binary cluster systems are transient structures that rapidly dissolve (or perhaps merge) under the combined influence of mutual and Galactic tidal forces \citep{priyatikanto16}. 

In summary, the existence of tidal binary clusters (TBCs) in the Milky Way remains unsupported by observational evidence. The available data instead favor a scenario in which binary clusters are predominantly primordial, young, and transient, with the oldest systems appearing to be no more than about $300~\text{Myr}$ old. This conclusion is consistent with the observational reassessment of old system candidates \citep{casado22a}, the dynamical instability demonstrated by numerical simulations,  the Primordial Group hypothesis \citep{casado22b}, and the failure of prominent candidates such as NGC~1605 to withstand detailed Gaia-based dynamical scrutiny.

Nevertheless, the two ambiguous pairs identified in \citet{casado22a} remain interesting targets for re-examination using improved Gaia data (e.g., radial velocities), homogeneous age determinations, refined membership assignments, and dynamical modeling capable of distinguishing primordial associations from true capture events. Additionally, extending kinematic studies to larger cluster systems could offer further insights into the survival timescales of cluster groups and the validity of the Primordial Group hypothesis.

\begin{acknowledgements}
Z.A. acknowledges financial support from the \.{I}lim Yayma Vakf{\i} (\.{I}YV).
\end{acknowledgements}

%\begin{thebibliography}{}

% --- Bibliography Setup ---
\bibliographystyle{aa}      
\bibliography{references}

\begin{thebibliography}{26}
\expandafter\ifx\csname natexlab\endcsname\relax\def\natexlab#1{#1}\fi

\bibitem[{{Al} {et~al.}(2026){Al}, {Karata{\c{s}}}, {Murillo}, {Bonatto},
  {G{\"u}ne{\c{s}}}, \& {{Akkaya}, I. and {{\"U}lgen}, Ey{\"u}p Kaan}}]{al2026}
{Al}, Z., {Karata{\c{s}}}, Y., {Murillo}, R.~M., {et~al.} 2026, Journal of
  Astrophysics and Astronomy, 47, 54

\bibitem[{{Anders} {et~al.}(2022){Anders}, {Castro-Ginard}, {Casado}, {Jordi},
  \& {Balaguer-N{\'u}{\~n}ez}}]{andres22}
{Anders}, F., {Castro-Ginard}, A., {Casado}, J., {Jordi}, C., \&
  {Balaguer-N{\'u}{\~n}ez}, L. 2022, Research Notes of the American
  Astronomical Society, 6, 58

\bibitem[{{Bhatia} \& {Hatzidimitriou}(1988)}]{Bhatia88}
{Bhatia}, R.~K. \& {Hatzidimitriou}, D. 1988, \mnras, 230, 215

\bibitem[{{Bhatia} {et~al.}(1991){Bhatia}, {Read}, {Hatzidimitriou}, \&
  {Tritton}}]{Bhatia91}
{Bhatia}, R.~K., {Read}, M.~A., {Hatzidimitriou}, D., \& {Tritton}, S. 1991,
  \aaps, 87, 335

\bibitem[{{Bressan} {et~al.}(2012){Bressan}, {Marigo}, {Girardi}, {Salasnich},
  {Dal Cero}, {Rubele}, \& {Nanni}}]{bressan12}
{Bressan}, A., {Marigo}, P., {Girardi}, L., {et~al.} 2012, \mnras, 427, 127

\bibitem[{{Camargo}(2021)}]{camargo21}
{Camargo}, D. 2021, \apj, 923, 21

\bibitem[{{Casado}(2022{\natexlab{a}})}]{casado22a}
{Casado}, J. 2022{\natexlab{a}}, Universe, 8, 368

\bibitem[{{Casado}(2022{\natexlab{b}})}]{casado22b}
{Casado}, J. 2022{\natexlab{b}}, Universe, 8, 113

\bibitem[{{Conrad} {et~al.}(2017){Conrad}, {Scholz}, {Kharchenko}, {Piskunov},
  {R{\"o}ser}, {Schilbach}, {de Jong}, {Schnurr}, {Steinmetz}, {Grebel},
  {Zwitter}, {Bienaym{\'e}}, {Bland-Hawthorn}, {Gibson}, {Gilmore},
  {Kordopatis}, {Kunder}, {Navarro}, {Parker}, {Reid}, {Seabroke}, {Siviero},
  {Watson}, \& {Wyse}}]{conrad17}
{Conrad}, C., {Scholz}, R.-D., {Kharchenko}, N.~V., {et~al.} 2017, \aap, 600,
  A106

\bibitem[{{Darma} {et~al.}(2021){Darma}, {Arifyanto}, \&
  {Kouwenhoven}}]{Darma21}
{Darma}, R., {Arifyanto}, M.~I., \& {Kouwenhoven}, M.~B.~N. 2021, \mnras, 506,
  4603

\bibitem[{{de La Fuente Marcos} \& {de La Fuente Marcos}(2009)}]{Marcos09}
{de La Fuente Marcos}, R. \& {de La Fuente Marcos}, C. 2009, \aap, 500, L13

\bibitem[{{de Oliveira} {et~al.}(2000){de Oliveira}, {Dutra}, {Bica}, \&
  {Dottori}}]{deoliveira2000}
{de Oliveira}, M.~R., {Dutra}, C.~M., {Bica}, E., \& {Dottori}, H. 2000, \aaps,
  146, 57

\bibitem[{{Dieball} {et~al.}(2002){Dieball}, {Mueller}, \&
  {Grebel}}]{diabel2002}
{Dieball}, A., {Mueller}, H., \& {Grebel}, E.~K. 2002, {VizieR Online Data
  Catalog: Binary and multiple clusters in the LMC (Dieball+, 2002)}, VizieR
  On-line Data Catalog: J/A+A/391/547. Originally published in:
  2002A\&A...391..547D

\bibitem[{{Gokmen} {et~al.}(2023){Gokmen}, {Eker}, {Yontan}, {Bilir}, {Ak},
  {Ak}, {Banks}, \& {Sarajedini}}]{gokmen23}
{Gokmen}, S., {Eker}, Z., {Yontan}, T., {et~al.} 2023, \aj, 166, 263

\bibitem[{{Groenewegen} \& {Salaris}(2003)}]{Groenewegen2003}
{Groenewegen}, M.~A.~T. \& {Salaris}, M. 2003, \aap, 410, 887

\bibitem[{{Hunt} \& {Reffert}(2024)}]{H24}
{Hunt}, E.~L. \& {Reffert}, S. 2024, {VizieR Online Data Catalog: Improving the
  open cluster census. III. (Hunt+, 2024)}, VizieR On-line Data Catalog:
  J/A+A/686/A42. Originally published in: 2024A\&A...686A..42H

\bibitem[{{Lada} \& {Lada}(2003)}]{lada2003}
{Lada}, C.~J. \& {Lada}, E.~A. 2003, \araa, 41, 57

\bibitem[{{Li} \& {Zhu}(2025)}]{li25}
{Li}, Z. \& {Zhu}, Z. 2025, \aap, 700, A280

\bibitem[{{Liu} {et~al.}(2025){Liu}, {Zhang}, {Zhong}, {Qin}, {Jiang}, \&
  {Chen}}]{liu25}
{Liu}, G., {Zhang}, Y., {Zhong}, J., {et~al.} 2025, \aap, 702, A48

\bibitem[{{Palma} {et~al.}(2025){Palma}, {Coenda}, {Baume}, \&
  {Feinstein}}]{palma25}
{Palma}, T., {Coenda}, V., {Baume}, G., \& {Feinstein}, C. 2025, \aap, 693,
  A218

\bibitem[{{Portegies Zwart} \& {Rusli}(2007)}]{portegies07}
{Portegies Zwart}, S.~F. \& {Rusli}, S.~P. 2007, \mnras, 374, 931

\bibitem[{{Priyatikanto} {et~al.}(2016){Priyatikanto}, {Kouwenhoven},
  {Arifyanto}, {Wulandari}, \& {Siregar}}]{priyatikanto16}
{Priyatikanto}, R., {Kouwenhoven}, M.~B.~N., {Arifyanto}, M.~I., {Wulandari},
  H.~R.~T., \& {Siregar}, S. 2016, \mnras, 457, 1339

\bibitem[{{Song} {et~al.}(2022){Song}, {Esamdin}, {Hu}, \& {Zhang}}]{song22}
{Song}, F., {Esamdin}, A., {Hu}, Q., \& {Zhang}, M. 2022, \aap, 666, A75

\bibitem[{{van den Bergh}(1996)}]{berg96}
{van den Bergh}, S. 1996, \apjl, 471, L31

\bibitem[{{Zhu} {et~al.}(2025){Zhu}, {Li}, \& {Zhang}}]{zhu25}
{Zhu}, Z., {Li}, Z., \& {Zhang}, S. 2025, \aap, 702, A259

\bibitem[{{Zhu} {et~al.}(2026){Zhu}, {Li}, \& {Zhang}}]{Zhu2026}
{Zhu}, Z., {Li}, Z., \& {Zhang}, S. 2026, \aap, 708, A87

\end{thebibliography}

%\end{thebibliography}

\begin{appendix}
	\section{Notes on individual systems}
	\label{sec:app_notes}
	
\nolinenumbers

\paragraph*{ASCC 108 and HSC 582}\hfill\break
	Although classified as an optical pair by \citet{palma25}, both clusters delineate a continuous, well-defined main sequence in the color-magnitude diagram (CMD). The member stars of both populations overlap seamlessly along the entire main sequence ($G \approx 11$--$18~\text{mag}$) and are accurately reproduced by a single theoretical isochrone. This high degree of photometric and evolutionary consistency renders a chance hyperbolic encounter or tidal capture highly unlikely, strongly supporting a primordial, coeval origin within the same giant molecular cloud (GMC).
	
	\paragraph*{CWNU 110 and NGC 2428}\hfill\break
	This system also incorporates HSC 1828, thus forming a triple cluster configuration \citep{palma25}. The dynamical assembly via tidal capture of three originally independent open clusters is exceptionally improbable. Given the complete overlap of member stars across the three CMDs and their consistent fit to a single isochrone, a primordial group origin provides a far more compelling explanation.
	
	\paragraph*{UBC 1320 and HSC 1585}\hfill\break
	While identified as a candidate TBC by \citet{liu25}, \citet{palma25} classified this system as an optical pair. Both components trace a tight and well-defined main sequence that is cleanly reproduced by a single isochrone. Their relatively young derived age ($\log(t/\text{yr}) = 7.9$) aligns with the absence of a prominent evolved stellar population in the CMD. Their tight spatial and photometric overlap makes a tidal capture event between independent systems highly improbable, reinforcing their nature as a coeval primordial pair.
	
	\paragraph*{Ruprecht 53 and HXHWL 65}\hfill\break
	\citet{palma25} classified this target as an optical pair. The stellar components of both clusters exhibit tight synchronicity throughout the main sequence and share identical turn-off features. The smooth continuity of member stars across both populations fitted to a single isochrone points directly to a shared birth origin rather than a dynamically forced capture or chance flyby.
	
	\paragraph*{NGC 2358 and Theia 366}\hfill\break
	This system forms a triple configuration together with HSC 1843, again raising significant dynamic objections to a tidal capture scenario. While NGC 2358 distinctly features an evolved red giant and red clump branch ($G \approx 9$--$10~\text{mag}$), all three components overlap along the main sequence and share a common turn-off point fitted by a single isochrone. This evolutionary consistency across both core and evolved phases further confirms a primordial origin from a single fragmenting cloud.
	
	\paragraph*{PHOC 40 and Theia 236}\hfill\break
	Designated as an optical pair by \citet{palma25}, both clusters contain a modest number of member stars and are categorized merely as moving groups by \citet{H24}. Because unbound moving groups cannot maintain internal gravitational binding, they cannot tidally capture or bind an orbiting companion cluster. Furthermore, the joint CMD demonstrates exceptional alignment across a broad magnitude range ($G \approx 9$--$18~\text{mag}$), with both systems fitted seamlessly by a single isochrone.
	
	\paragraph*{Trumpler 23 and Teutsch 247}\hfill\break
	Despite being classified as an optical pair by \citet{palma25}, and exhibiting a slightly higher degree of scatter than other candidates, member stars in this system share a distinct main sequence in the joint CMD with a clear turn-off region around $G \approx 13.5~\text{mag}$. The compatibility of both populations with a single isochrone supports a common birth cloud over a dynamical capture or hyperbolic encounter.
	
	\paragraph*{Theia 2151 and UBC 502}\hfill\break
	Although categorized as a merely optical pair by \citet{palma25}, the joint CMD reveals a unified, cohesive main sequence and sub-giant/turn-off morphology. This elevated degree of photometric and evolutionary harmony strongly favors formation within a common birth cloud over accidental tidal pairing.
	
	\paragraph*{Theia 4284 and Theia 4981}\hfill\break
	Also reported as an optical pair by \citet{palma25}, both member populations blend seamlessly into a single main-sequence track well reproduced by our best-fit isochrone. This morphological and evolutionary coherence confirms a primordial association originating from the same parent cloud.
	
\end{appendix}

\end{document}